\documentclass[journal, twocolumn]{IEEEtran}

\usepackage{amsmath,amssymb,graphicx,epsfig,fancyhdr,amsthm,tabulary,epstopdf,stmaryrd}
\usepackage{hyperref}
\usepackage{cite}
\usepackage{multirow}
\usepackage{mpmulti}
\usepackage{float}
\usepackage{amsfonts}
\usepackage{subfigure}
\usepackage{footnote}
\makesavenoteenv{tabular}
\usepackage{tabulary,bigstrut,array,multirow,booktabs}%
\usepackage{tikz}
\usetikzlibrary{calc}
\usepackage{pdfsync}
\usepackage{bm}
\usepackage{balance}

\theoremstyle{remark}

\theoremstyle{definition}

\begin{document}

 	\title{Deep Autoencoder-Based Constellation Design \\ in Multiple Access Channels} 
 	

 	\author{%
 		\IEEEauthorblockN{Stepan Gorelenkov and  Mojtaba Vaezi}\\
 		\IEEEauthorblockA{Department of Electrical and Computer Engineering\\
 			Villanova University,
 			Villanova, PA 19085\\
 			Emails: \{sgorelen, mvaezi\}@villanova.edu}
 \thanks{This work was supported by the NSF under Grant ECCS-2301778.}	}

\maketitle


\begin{abstract} 
	In multiple access channels (MAC), multiple users share a transmission medium to communicate with a common receiver. Traditional constellations like quadrature amplitude modulation are optimized for point-to-point systems and lack mechanisms to mitigate inter-user interference, leading to suboptimal performance in MAC environments. To address this, we propose a novel framework for constellation design in MAC that employs deep autoencoder (DAE)-based communication systems. This approach intelligently creates flexible constellations aware of inter-user interference, reducing symbol error rate and enhancing the constellation-constrained sum capacity of the channel. Comparisons against analytically derived constellations demonstrate that DAE-designed constellations consistently perform best or equal to the best across various system parameters. Furthermore, we apply the DAE to scenarios where no analytical solutions have been developed, such as with more than two users, demonstrating the adaptability of the model.
\end{abstract}

\section{Introduction}

A multiple access channel (MAC) allows multiple users to simultaneously transmit over a shared medium to a common receiver \cite{cover2006elements,el2011network,liao1972multiple,ahlswede1971two}. The capacity region of a MAC is achieved via non-orthogonal transmission, where users transmit simultaneously within the same frequency band. For a two-user Gaussian MAC, the capacity region forms a pentagon and is obtained by inputs that are continuous and Gaussian
distributed whose length tends to infinity \cite{cover2006elements,el2011network}. Although Gaussian-distributed inputs achieve capacity and offer theoretical insights, they do not reflect achievable rates under practical constraints, such as with finite constellations like quadrature amplitude modulation (QAM).

Motivated by this, constellation-constrained (CC) achievable rates for the two-user Gaussian MAC under both orthogonal and non-orthogonal transmission schemes were evaluated in \cite{harshan2011two}. The study also proposed optimizing the rotation angle between input constellations to maximize the CC-achievable sum rate by reducing the symbol overlap at the receiver caused by the superposition of individual constellation symbols.
Similar concepts have been developed for other variants of MAC such as cognitive MAC \cite{zhang2009cognitive,liu2011fading,liu2019minimum}.  MAC has also recently been explored under the name of uplink non-orthogonal multiple access (NOMA) \cite{vaezi2019noma,liu2019minimum} in the communication society.

In finite-alphabet MAC, the primary challenge is designing symbol sets for each user to ensure accurate recovery of each user's message at the receiver after a superposition of individual constellation symbols. Traditional constellations, such as QAM, are optimized for point-to-point systems with predefined symbols and lack mechanisms to mitigate inter-user interference. This inability to respond to interference is an obstacle to improving the bit-error rate and spectral efficiency of MAC where inter-user interference is present. 

Recently, deep autoencoders (DAEs) have been used to enhance communication system performance by enabling end-to-end design \cite{Shea2017}. This idea has since been extended to various applications, including constellation design for interference-limited  channels \cite{wu2020deep, zhang2024deep, bo2024joint, Alberge,Jafarkhani2024noma}, as well as interference cancellation in joint source channel coding and MAC \cite{yilmaz2023distributed, shlezinger2020deepsic}. However, to date, there has been no DAE-based constellation design specifically for the MAC.

Thus,  we propose a DAE-based framework for constellation design in MAC that intelligently creates flexible constellations aware of inter-user interference. This allows constellations to adapt to interference patterns, thus reducing the symbol error rates of the users and enhancing the CC-achievable sum rate of the MAC. The approach leverages DAEs to jointly optimize the encoding and decoding processes, leading to more robust and efficient communication strategies.
The constellations designed by the DAE are compared against several analytically derived constellations. While the best-performing analytical models exhibited poor performance under some scenarios, the DAE-generated constellations consistently achieved the best or near-best performance in all considered system parameters.


\section{System Model}

\setcounter{footnote}{0}  

\begin{figure}
	\centering
	\includegraphics[width=1\linewidth]{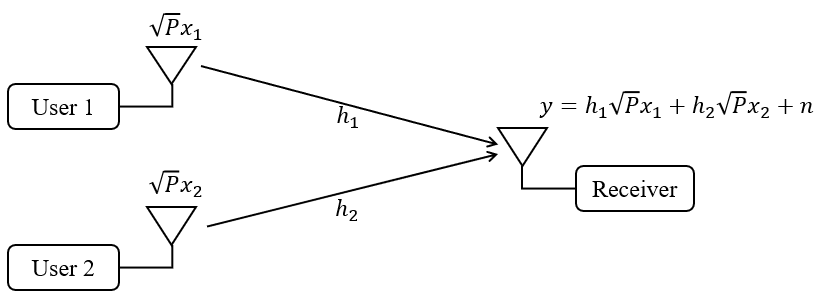}
	\caption{Two-user MAC model with the assumed additive Gaussian noise.}
	\label{fig:sys-model}
\end{figure}

This work primarily focuses on the two-user MAC, \footnote{However, the DAE model can be generalized to accommodate any number of users as will be demonstrated in Section~\ref{sec:Extension}.} shown in Fig.~\ref{fig:sys-model}, in which each user transmits symbols from their respective constellations, \(\mathcal{S}_1\) and \(\mathcal{S}_2\), containing \(|\mathcal{S}_1|\) and \(|\mathcal{S}_2|\) symbols, respectively, with \(x_i \in \mathcal{S}_i\) for \(i \in \{1,2\}\) \cite{harshan2011two}. The transmitted symbols are subject to an average power constraint \(P_i\), and the constellations are normalized so that \(\mathbb{E}[|x_i|^2] = 1\). The received signal is then given by
\begin{equation}
y = h_1 \sqrt{P_1} x_1 + h_2 \sqrt{P_2} x_2 + n,
\end{equation}
where \(h_1\) and \(h_2\) are the channel gains, and $n$ denotes additive white complex Gaussian  noise, whose real and imaginary elements are sampled from $\mathcal{N}(0,\frac{N_0}{2})$ with $N_0$ being the total noise power \cite{Grey_Uplink}.

Assuming a non-power sharing mode with equal maximum power constraints for each user ($P_1 = P_2 = P$), the received power from each user becomes $h_1^2 P$ and $h_2^2 P$, respectively. The average received power is then $P_{\text{av}} = \frac{h_1^2 P + h_2^2 P}{2}$. Defining $\alpha$ as the ratio of power received from User 2 to the average received power, and normalizing $P_{\rm{av}}$ to 1 for simplicity, the system equation can be rewritten as \cite{Grey_Uplink,liu2019minimum} 
\begin{equation}
y = \sqrt{2 - \alpha} x_1 + \sqrt{\alpha} x_2 + n.
\end{equation}

This formulation facilitates the analysis of constellation performance across different signal-to-noise ratios (SNRs).


\section{DAE Model and Training}
The design problem for this system is to find the optimal constellations for each user given a set of system parameters $\{\alpha,k_1,k_2\}$ where $k_i$ is the number of bits sent per symbol for user $i$. This results in the size of User-$i$'s constellation $|\mathcal S_i|=2^{k_i}$. Given the received symbol, the receiver needs to recover the bits sent by each user. 
The DAE model is adapted from \cite{Alberge} to fit the MAC system model. 
\subsection{Transmitters}
\begin{figure}
	\centering
	\includegraphics[width=0.8\linewidth]{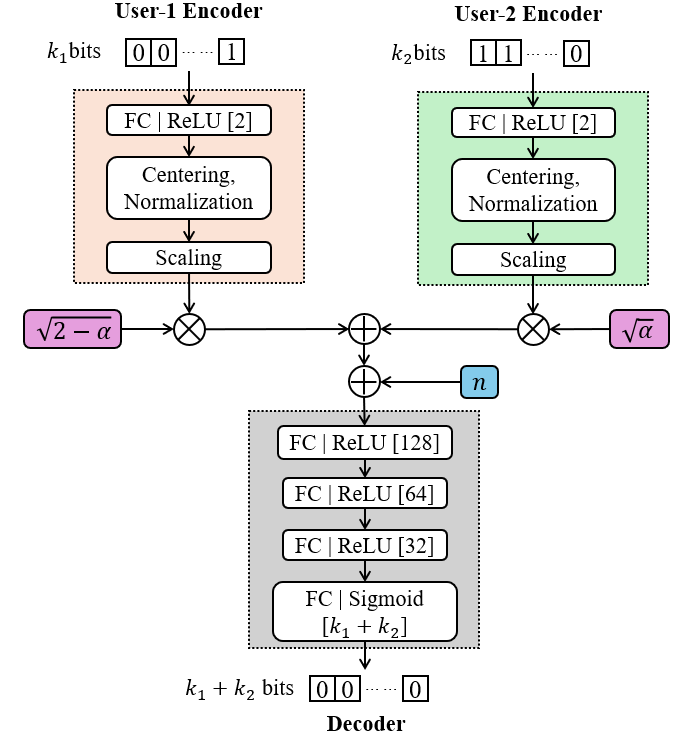}
	\caption{Model describing DAE layers. FC denotes a fully connected layer, with the activation function and the corresponding number of neurons indicated in brackets.}
	\label{fig:DAE-model}
\end{figure}

There are two transmitters in the case considered, labeled \textit{User-1 Encoder and User-2 Encoder} in Fig. \ref{fig:DAE-model}. Transmitter~\(i\) encodes a \(k_i\)-bit sequence into a unique in-phase and quadrature (I/Q) symbol, subject to the system's power constraints.
Encoding is performed using the bit sequence directly \cite{zhang2022svd}, rather than one-hot encoding \cite{Shea2017}, as it has been shown to provide better performance in such applications and naturally aligns with how users present data as sequences of bits.
Bit information is encoded using a single fully connected (FC) layer with two neurons, representing the I and Q components of the symbol, followed by ReLU activation.
This results in a very simple encoder, making it practically attractive.
The next two layers enforce the power constraints by centering and then normalizing the symbol. This is done by enforcing a mean of zero and an average power of 1 across the entire test batch. Each test batch consists of $\mathrm{Num Const}$ copies of every possible permutation of bits. Thus, a total of $\mathrm{Num Const}$ identical constellations are created by the transmitter in each batch. After this, scaling is applied by dividing each dimension by a learned parameter, which is constrained to the range \([1, \infty)\) using the ReLU function.
This allows for constellations to have lower than maximum power if needed. Then, as in the channel the symbol from each user are scaled according to the $\alpha$ for the channel.

\subsection{Receiver}
Afterward, the users' symbols are summed to produce the aggregate symbol, to which Gaussian noise with power $\frac{N_0}{2}$ per I and Q dimension is added, resulting in the received symbol.
This symbol is input to the receiver, which consists of three FC layers with 128, 64, and 32 neurons, each followed by ReLU activation, and a final layer with \(k_1+k_2\) neurons and sigmoid activation, as shown in Fig.~\ref{fig:DAE-model}.
The number of neurons is relatively large to improve the decoding of received symbols, as smaller models exhibited poorer performance, suggesting that the larger size aids learning at the transmitter. Nevertheless, the computational complexity of the receiver remains relatively low, as it primarily involves a few matrix multiplications. The receiver outputs a vector $z$ of length $k_1+k_2$, where each element $z_j$ represents the estimated probability that the transmitted bit $b_j$ is one.

\subsection{Training Parameters}
In \cite{Stacked_Denoising}, it is shown that, minimizing the cross-entropy between the input bits and their output probabilities is equivalent to maximizing a lower bound on their mutual information.
Thus, the cross entropy was used as a loss function. The model was trained with $\rm{NumConst}=2^{11}$, and up to $2^{15}$ epochs when the number of bits and the number of users were high, and was terminated if no improvement in the loss was observed over the previous 4000 epochs. For each constellation designed, the model was trained with a constant SNR. In each scenario, the model was trained multiple times using various fixed SNR values, typically 10, 13, and 16~dB. For the results presented, the best-performing model from these runs was selected for each case. The Adam optimizer was used with the default parameters provided by Keras. In Section~\ref{sec:Extension}, the training SNRs were higher, ranging from 20 to 34~dB.

\section{Results}

The performance of the DAE-designed constellation is compared with optimized traditional constellations using the CC-achievable sum rate, referred to as the CC sum-capacity in \cite{harshan2011two}, and defined by
\begin{equation}
I(\sqrt{2-\alpha} x_1,\sqrt{\alpha}x_2;y).
\end{equation}
This quantity is defined as the mutual information between the input and output of a Gaussian channel, where (i) the input constellation is finite and (ii) the constellation symbols are uniformly distributed. A general formula for the CC-achievable  rates of two-user constellations, with or without rotation, was derived in \cite{harshan2011two}.

\subsection{Constellations Used for Comparison}
We use the CC-achievable sum rate metric to compare the DAE-designed constellation against three analytically derived constellations, as listed below.

\begin{itemize}
	\item 
	In \cite{harshan2011two}, optimal rotation angles for equal-powered quadrature phase shift keying (QPSK) constellations were found for both users to maximize the CC sum rate in the case of $k_1 = k_2 = 2$. Since the rotation angles were not explicitly provided for  QPSK and BPSK constellations ($k_1 = 2$ and $k_2 = 1$), we applied their method to compute the optimal rotation for this case.
	It should be noted that, they also compare this against the case where each constellation is a pulse-amplitude modulation (PAM), rotated in a way that each user's message is orthogonal, leading to 4-PAM having superior performance in both equal power cases compared to optimally rotated QPSK. 
	\item In \cite{QPSKRot}, the optimal rotation angles for two QPSK constellations are derived by maximizing the minimum distance (MD) between symbols in the general case for any $\alpha$.
	
	\item The analytical constellation designed in \cite{liu2019minimum} optimizes for MD using a parallelogram structure for both users, and demonstrates better symbol error rate (SER) performance than the design in \cite{QPSKRot} in the case of $\alpha = 0.8$ for both $k_1 = k_2 = 2$ and $k_1 = 2, k_2 = 1$.
	
\end{itemize}

\subsection{Case with $\alpha=1$}
\begin{figure}[h]
	\centering
	\includegraphics[width=.9\linewidth]{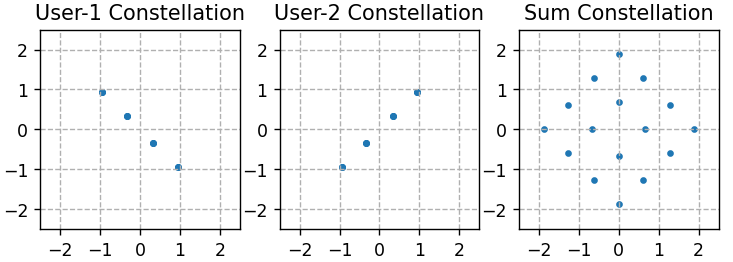}
	\includegraphics[width=.9\linewidth]{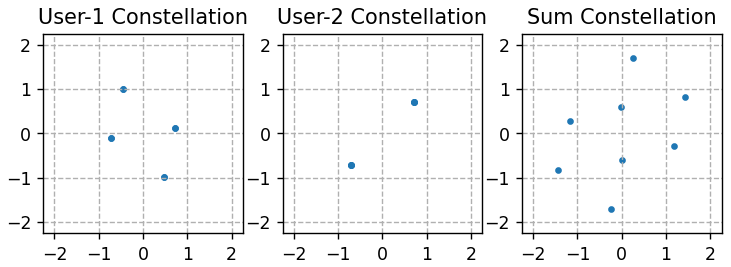}
	\caption{DAE-designed constellations with (top) $k_1 = k_2 = 2$ and (bottom) $k_1 = 2$, $k_2 = 1$, with $\alpha=1$.}
	\label{fig:Const a_1}
\end{figure}

First, the case with $\alpha=1$ is investigated which corresponds to case with $h_1=h_2$. The DAE was used to design constellations for both $k_1=k_2=2$ and $k_1=2,k_2=1$, shown in Fig.~\ref{fig:Const a_1}. Here, the individual constellations for each user are shown to the left, and the sum constellation is shown on the right. The sum constellation is the set of all possible symbols that can be received. It can be found by superimposing the constellations of each user.

\begin{figure}[h]
	\centering
	\includegraphics[width=.859\linewidth]{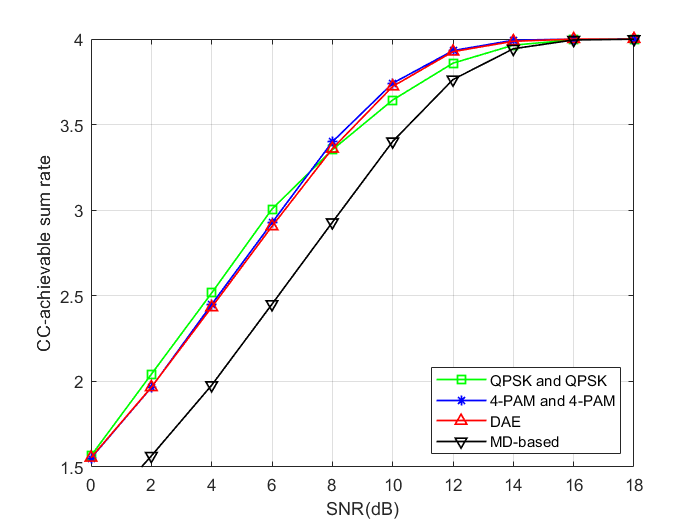}
	\includegraphics[width=.859\linewidth]{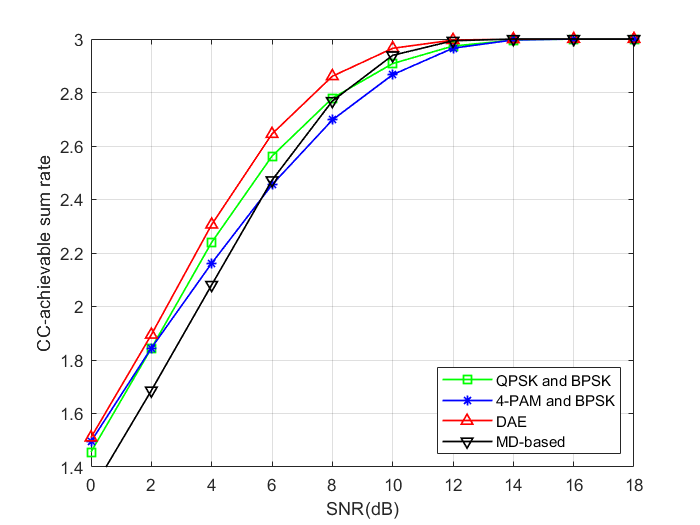}
	\caption{CC-achievable sum rate for $\alpha = 1$, comparing the DAE-designed constellation with various approaches for (top) $k_1 = k_2 = 2$ and (bottom) $k_1 = 2$, $k_2 = 1$.}
	\label{fig:CC a_1}
\end{figure}

We note that in the case of $k_1=k_2=2$, the resulting constellation is a rotated 4-PAM for each user. In Fig. \ref{fig:CC a_1}, the  CC-achievable sum rate of these constellations are compared to traditional constellations with optimal rotation angles chosen from \cite{harshan2011two}, and the MD-based method applied to $\alpha=1$.

It is seen that the CC-achievable  sum rate is the same for 4-PAM and the DAE as both user's constellations are orthogonal. They both approach the maximum sum rate quicker than optimally rotated QPSK and MD-based. We can see that when $k_2=1$, in both the 4-PAM and QPSK case, one of the user's constellation becomes BPSK, resulting a poorer performance than the others. In this case, the MD method seems to approach the sum capacity limit at the same SNR as the DAE, but has a steeper drop off as SNR decreases.
\begin{figure}[t]
	\centering
	\includegraphics[width=.79\linewidth]{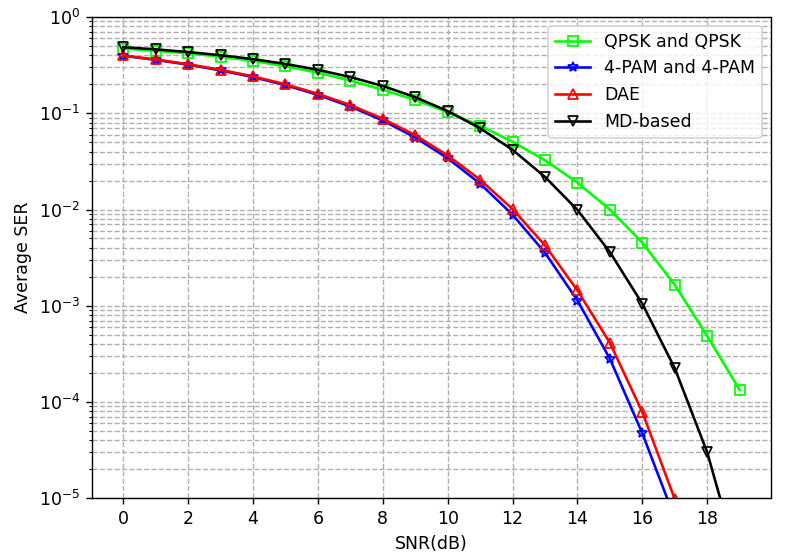}
	\includegraphics[width=.79\linewidth]{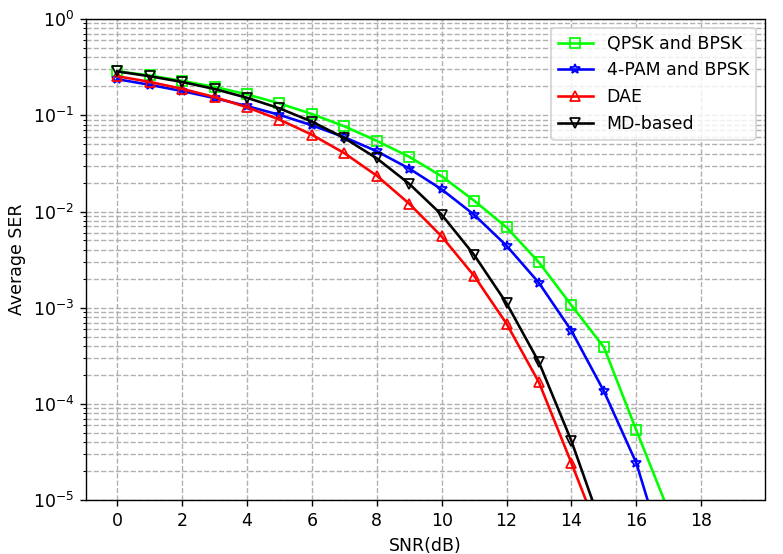}
	\caption{Symbol error rate comparison for different constellations with $\alpha = 1$ and (top) $k_1 = k_2 = 2$, (bottom) $k_1 = 2$, $k_2 = 1$.}
	\label{fig:SER a_1}
\end{figure}
In Fig.~\ref{fig:SER a_1}, the SERs of the constellations are compared. The SER was calculated through simulation with using a maximum likelihood receiver for the analytical constellations, and the DAE decoder section for the DAE constellation. In both cases, DAE performs with the best, or nearly best SER. In the $k_1=k_2=2$ case, the 4-PAM pair constellation noticeably outperforms the next-best MD-based constellation. Conversely, when $k_2=1$, the MD-based constellation outperforms the 4-PAM. This demonstrates the adaptability of the DAE to find well-performing constellations across various system parameters.

\subsection{Case with $\alpha$ = 0.8}
\begin{figure}[h]
	\centering
	\includegraphics[width=0.82\linewidth]{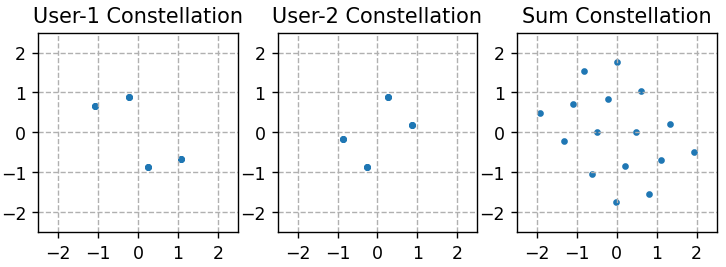}
	\includegraphics[width=0.82\linewidth]{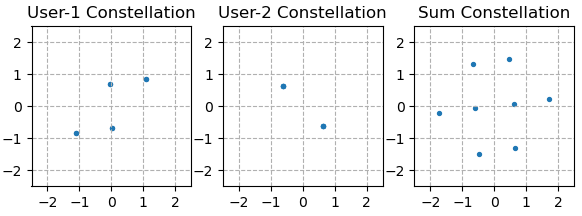}
	\caption{DAE-designed constellations for (top) $k_1 = k_2 = 2$ and (bottom) $k_1 = 2$, $k_2 = 1$, with $\alpha = 0.8$.}
	\label{fig:Const_a08}
\end{figure}

Next, we train a new model for $\alpha = 0.8$. We compare its performance against the same analytical constellation design models, with the modification that the optimal QPSK rotation angles are now calculated based on \cite{QPSKRot}. The DAE-designed constellations are shown in Fig.~\ref{fig:Const_a08}.

\begin{figure}[th]
	\centering
	\includegraphics[width=.79\linewidth]{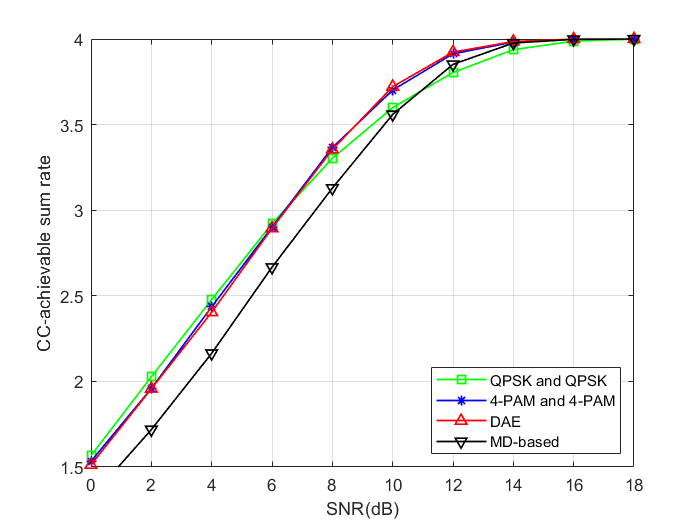}
	\includegraphics[width=.79\linewidth]{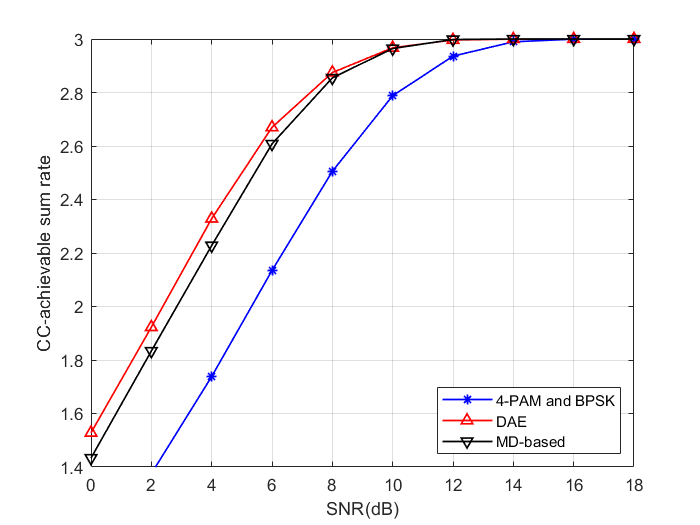}
	\caption{CC-achievable  sum rates of constellations considered with $\alpha=0.8$ and (top) $k_2=2$, (bottom) $k_2=1$.}
	\label{fig:CC_a08}
\end{figure}

Note that the shape of the constellation is different from that with $\alpha = 1$, which shows that for different $\alpha$ values, the optimal sum constellations may change. This highlights the advantage of using a machine learning architecture, as it can discover complex statistical relationships between the available degrees of freedom and the optimal constellation.

\begin{figure}[h]
	\centering
	\includegraphics[width=.82\linewidth]{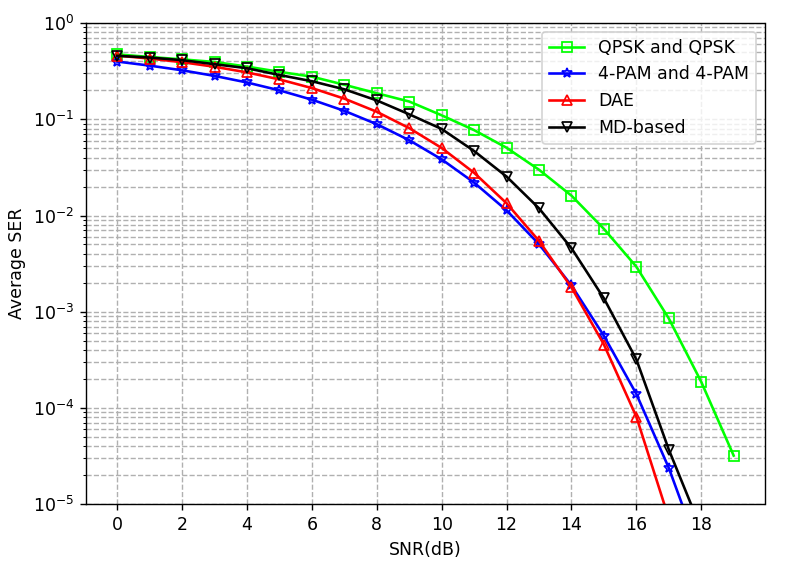}
	\includegraphics[width=.82\linewidth]{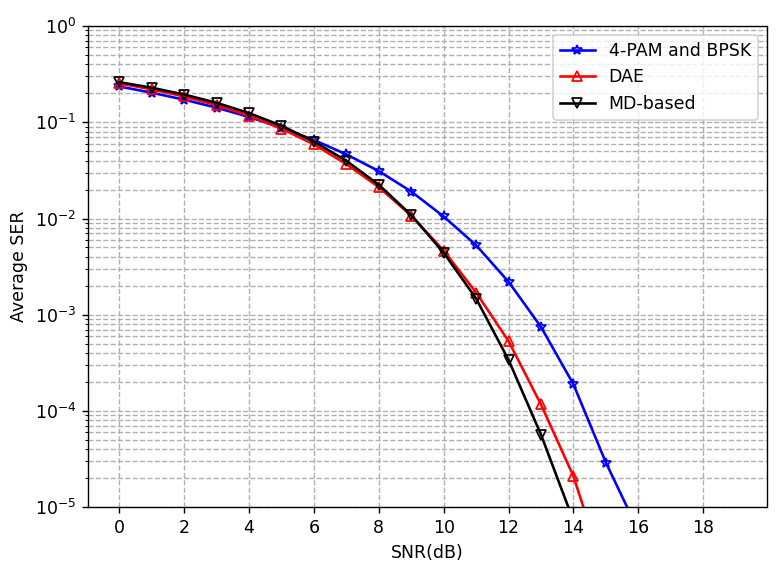}
	\caption{Symbol error rate comparing constellations with $\alpha=0.8$ and (top) $k_1=k_2=2$, (bottom) $k_1=2,k_2=1$.}
	\label{fig:SER a_08}
\end{figure}

\begin{figure}[th]
	\centering
	\includegraphics[width=.99\linewidth]{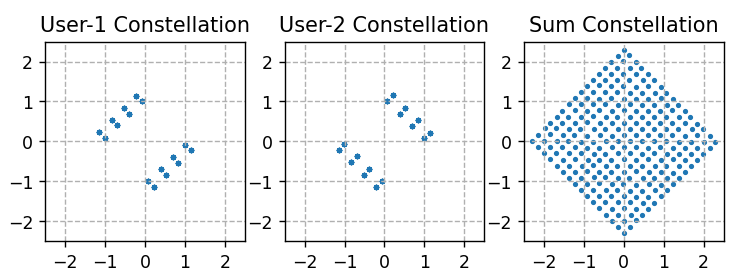}
	\caption{DAE designed constellation for $k_1=k_2=4$ and $\alpha=1$.}
	\label{fig:k=4}
\end{figure}
Figure~\ref{fig:CC_a08} shows the CC-achievable sum rate of all constellations with \( \alpha = 0.8 \), demonstrating that the DAE outperforms or matches the best-performing analytical constellations. While the MD-based constellation performs well, it fails to adapt effectively across all scenarios.
We also compare the SERs in Fig.~\ref{fig:SER a_08}, which shows that with $\alpha=0.8$, the DAE-designed constellation matches the SER performance of the best-performing analytical constellations.
The best-performing analytical constellation in terms of SER differs between the cases of $k_2=2$ and $k_2=1$, illustrating that the DAE is the only method adaptable to changing system parameters and capable of discovering well-performing constellations for each scenario. In addition, all previous SER results demonstrate the receiver’s ability to correctly identify decision regions.

\section{Extension to Three Users and Higher-Order Constellations}
\label{sec:Extension}

We design constellations for two additional cases to further demonstrate the versatility of the DAE. First, we consider the case with $\alpha = 1$, $k_1 = k_2 = 4$, and plot the corresponding constellation in Fig.~\ref{fig:k=4}. 
More importantly, we train a DAE for the three-user MAC.

In extending to channels with more than two users, we redefine the channel parameter $\alpha$, as a single value is no longer sufficient to characterize all channels. Instead, we define a vector $\bm{\alpha} = [\alpha_1, \alpha_2, \ldots, \alpha_K]$, where $\alpha_i = \frac{h_i^2 P}{P_{\mathrm{av}}}$ represents the ratio of the power received from the user $i$ to the average power received, and $K$ is the number of users. 

\begin{figure}[t]
	\centering
	\includegraphics[width=0.892\linewidth]{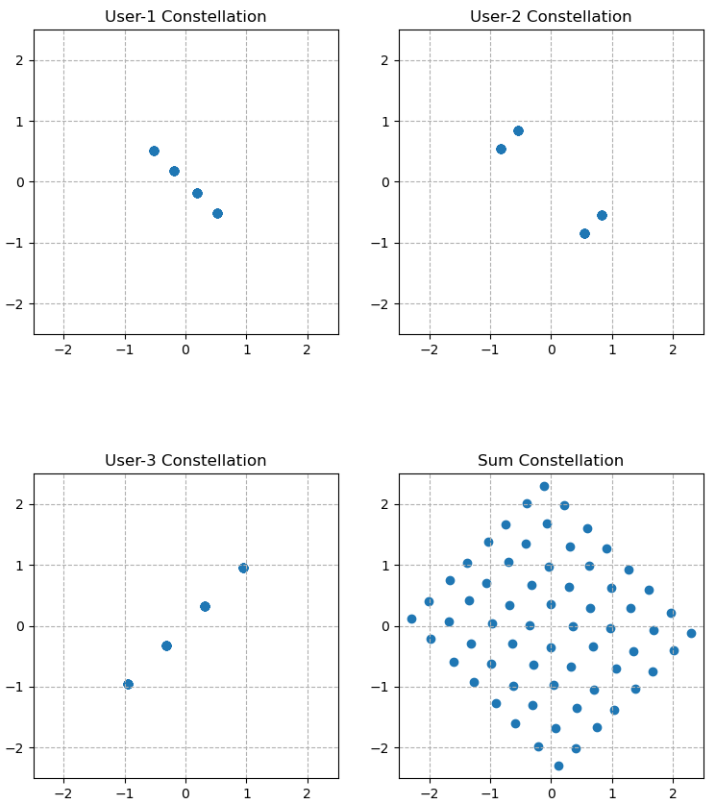}
	\caption{DAE-designed constellation for a three-user MAC with $k_1 = k_2 = k_3 = 2$ and $\bm{\alpha} = [1, 1, 1]$.}
	\label{fig:M=3}
\end{figure}

The case considered in this section is $\bm{\alpha} = [1, 1, 1]$, with $k_1 = k_2 = k_3 = 2$, where a third user is added by using an additional encoder before the summation. The corresponding constellation is depicted in Fig.~\ref{fig:M=3}.

Although we were unable to find optimally designed analytical constellations for these two cases to use as a benchmark,
the designed constellations appear promising. Since the SER depends on the sum constellation, we can use its structure as an indication of expected performance. As the sum constellations resemble 256-QAM and 64-QAM, respectively, we expect that the constellations designed for each user will yield good performance.

\section{Conclusion}
We have demonstrated that a DAE can design constellations that either outperform or match the performance of analytically designed constellations for the MAC. Among all the methods considered, the DAE was the only one capable of consistently producing the best-performing constellation in all cases. In addition to designing higher-order constellations, this highlights the adaptability of the DAE.

The main contribution of this work is to validate that a machine learning model can find near-optimal constellations. However, the designed DAE is limited in that a new model must be trained for each set of system parameters ${K, \alpha, k_1, k_2}$.  This provides an avenue for future research toward developing a model trained across a wide range of system parameters capable of outputting an optimal constellation given the system parameters as input.

\newpage
\balance

\begin{thebibliography}{10}
	\providecommand{\url}[1]{#1}
	\csname url@samestyle\endcsname
	\providecommand{\newblock}{\relax}
	\providecommand{\bibinfo}[2]{#2}
	\providecommand{\BIBentrySTDinterwordspacing}{\spaceskip=0pt\relax}
	\providecommand{\BIBentryALTinterwordstretchfactor}{4}
	\providecommand{\BIBentryALTinterwordspacing}{\spaceskip=\fontdimen2\font plus
		\BIBentryALTinterwordstretchfactor\fontdimen3\font minus
		\fontdimen4\font\relax}
	\providecommand{\BIBforeignlanguage}[2]{{%
			\expandafter\ifx\csname l@#1\endcsname\relax
			\typeout{** WARNING: IEEEtran.bst: No hyphenation pattern has been}%
			\typeout{** loaded for the language `#1'. Using the pattern for}%
			\typeout{** the default language instead.}%
			\else
			\language=\csname l@#1\endcsname
			\fi
			#2}}
	\providecommand{\BIBdecl}{\relax}
	\BIBdecl
	
	\bibitem{cover2006elements}
	T.~M. Cover, \emph{Elements of Information Theory}.\hskip 1em plus 0.5em minus
	0.4em\relax John Wiley \& Sons, 2006.
	
	\bibitem{el2011network}
	A.~El~Gamal and Y.-H. Kim, \emph{Network Information Theory}.\hskip 1em plus
	0.5em minus 0.4em\relax Cambridge University Press, 2011.
	
	\bibitem{liao1972multiple}
	H.~Liao, ``Multiple access channels,'' \emph{Ph.D. dissertation, Dept. Elec.
		Eng., Univ. of Hawaii}, 1972.
	
	\bibitem{ahlswede1971two}
	R.~Ahlswede, ``On two-way communication channels and a problem by
	zarankiewicz,'' \emph{Problems of Control and Information Theory}, pp.
	23--37, 1971.
	
	\bibitem{harshan2011two}
	J.~Harshan and B.~S. Rajan, ``On two-user {Gaussian} multiple access channels
	with finite input constellations,'' \emph{IEEE Transactions on Information
		Theory}, vol.~57, no.~3, pp. 1299--1327, 2011.
	
	\bibitem{zhang2009cognitive}
	L.~Zhang, Y.~Xin, Y.-C. Liang, and H.~V. Poor, ``Cognitive multiple access
	channels: optimal power allocation for weighted sum rate maximization,''
	\emph{IEEE Transactions on Communications}, vol.~57, no.~9, pp. 2754--2762,
	2009.
	
	\bibitem{liu2011fading}
	R.~Liu, Y.~Liang, and H.~V. Poor, ``Fading cognitive multiple-access channels
	with confidential messages,'' \emph{IEEE Transactions on Information Theory},
	vol.~57, no.~8, pp. 4992--5005, 2011.
	
	\bibitem{liu2019minimum}
	H.-P. Liu, S.-L. Shieh, C.-H. Lin, and P.-N. Chen, ``{A minimum distance
		criterion based constellation design for uplink NOMA},'' in \emph{Proc. IEEE
		Vehicular Technology Conference (VTC2019-Fall)}, 2019, pp. 1--5.
	
	\bibitem{vaezi2019noma}
	M.~Vaezi and H.~Vincent~Poor, ``{NOMA: An information-theoretic perspective},''
	\emph{{Multiple Access Techniques for 5G Wireless Networks and Beyond}}, pp.
	167--193, 2019.
	
	\bibitem{Shea2017}
	T.~O’Shea and J.~Hoydis, ``An introduction to deep learning for the physical
	layer,'' \emph{IEEE Transactions on Cognitive Communications and Networking},
	vol.~3, no.~4, pp. 563--575, 2017.
	
	\bibitem{wu2020deep}
	D.~Wu, M.~Nekovee, and Y.~Wang, ``Deep learning-based autoencoder for m-user
	wireless interference channel physical layer design,'' \emph{IEEE Access},
	vol.~8, pp. 174\,679--174\,691, 2020.
	
	\bibitem{zhang2024deep}
	X.~Zhang and M.~Vaezi, ``Deep autoencoder-based {Z}-interference channels with
	perfect and imperfect {CSI},'' \emph{IEEE Transactions on Communications},
	vol.~72, no.~2, pp. 861--873, 2024.
	
	\bibitem{bo2024joint}
	Y.~Bo, Y.~Duan, S.~Shao, and M.~Tao, ``Joint coding-modulation for digital
	semantic communications via variational autoencoder,'' \emph{IEEE
		Transactions on Communications}, 2024.
	
	\bibitem{Alberge}
	F.~Alberge, ``Constellation design with deep learning for downlink
	non-orthogonal multiple access,'' in \emph{Proc. IEEE International Symposium
		on Personal, Indoor and Mobile Radio Communications (PIMRC)}, 2018, pp. 1--5.
	
	\bibitem{Jafarkhani2024noma}
	H.~Jafarkhani, H.~Maleki, and M.~Vaezi, ``Modulation and coding for {NOMA} and
	{RSMA},'' \emph{Proceedings of the IEEE}, vol. 112, no.~9, pp. 1--35, 2024.
	
	\bibitem{yilmaz2023distributed}
	S.~F. Yilmaz, C.~Karamanl{\i}, and D.~G{\"u}nd{\"u}z, ``Distributed deep joint
	source-channel coding over a multiple access channel,'' in \emph{Proc. IEEE
		International Conference on Communications (ICC)}, 2023, pp. 1400--1405.
	
	\bibitem{shlezinger2020deepsic}
	N.~Shlezinger, R.~Fu, and Y.~C. Eldar, ``{DeepSIC: Deep soft interference
		cancellation for multiuser MIMO detection},'' \emph{IEEE Transactions on
		Wireless Communications}, vol.~20, no.~2, pp. 1349--1362, 2020.
	
	\bibitem{Grey_Uplink}
	S.-L. Shieh, C.-H. Lin, Y.-C. Huang, and C.-L. Wang, ``On {Gray} labeling for
	downlink non-orthogonal multiple access without {SIC},'' \emph{IEEE
		Communications Letters}, vol.~20, no.~9, pp. 1721--1724, 2016.
	
	\bibitem{zhang2022svd}
	X.~Zhang, M.~Vaezi, and T.~J. O’Shea, ``{SVD-embedded deep autoencoder for
		MIMO communications},'' in \emph{Proc. IEEE International Conference on
		Communications (ICC)}, 2022, pp. 5190--5195.
	
	\bibitem{Stacked_Denoising}
	\BIBentryALTinterwordspacing
	P.~Vincent, H.~Larochelle, I.~Lajoie, Y.~Bengio, and P.-A. Manzagol, ``Stacked
	denoising autoencoders: Learning useful representations in a deep network
	with a local denoising criterion,'' \emph{Journal of Machine Learning
		Research}, vol.~11, no. 110, pp. 3371--3408, 2010. [Online]. Available:
	\url{http://jmlr.org/papers/v11/vincent10a.html}
	\BIBentrySTDinterwordspacing
	
	\bibitem{QPSKRot}
	C.-H. Lin, S.-L. Shieh, T.-C. Chi, and P.-N. Chen, ``Optimal
	inter-constellation rotation based on minimum distance criterion for uplink
	{NOMA},'' \emph{IEEE Transactions on Vehicular Technology}, vol.~68, no.~1,
	pp. 525--539, 2019.
	
\end{thebibliography}


\end{document}